\documentstyle[preprint,aps,prl]{revtex} 
\tightenlines
\begin{document}
\ifx\undefined\psfig\def\psfig#1{    }\else\fi     \ifpreprintsty\else
\twocolumn[\hsize\textwidth
\columnwidth\hsize\csname@twocolumnfalse\endcsname       \fi    \draft
\preprint{  }  \title  {Exact   exchange-correlation potential for   a
time-dependent  two   electron   system}  \author  {Irene D'Amico  and
Giovanni Vignale}   \address{Department  of Physics,   University   of
Missouri, Columbia, Missouri 65211} \date{\today} \maketitle 

\begin{abstract}
We  obtain an  exact  solution  of  the  time-dependent  Schr\"odinger
equation for a two-electron system confined to a plane by an isotropic
parabolic potential whose curvature is periodically modulated in time.
From this solution we compute  the {\it exact} time-dependent exchange
correlation potential $v_{xc}$ which enters  the Kohn-Sham equation of
time-dependent density functional theory.  Our exact result provides a
benchmark against which various  approximate forms for $v_{xc}$ can be
compared.  Finally  $v_{xc}$ is separated  in an adiabatic and  a pure
dynamical   part  and it  is shown   that,  for  the particular system
studied, the dynamical part is negligible. 
\end{abstract}  
 \ifpreprintsty\else\vskip1pc]\fi \narrowtext

\section{Introduction}
The Time Dependent  Density Functional Theory (TDFT) \cite{RG,GDP,DDV}
maps an interacting time-dependent  N-electron system, described  by a
Hamiltonian of the form 
\begin {equation}
H  =  \sum_i  \frac{p_i^2}{2m}+\sum_{i<j}V({\bf  r}_i-{\bf   r}_j)   +
\sum_{i}v_{ext}({\bf    r}_i,t)    \end   {equation}     with    ${\bf
p}_i=-i\hbar\nabla_i$  the momentum  operator of the  single particle,
$V({\bf  r}_i-{\bf r}_j)$    the   two-particle interaction  potential
($V({\bf  r}_i-{\bf  r}_j)=e^2/|{\bf   r}_i-{\bf   r}_j|$  for Coulomb
interaction)  and $v_{ext}({\bf  r}_i,t)$  the time-dependent external
potential,  to a  non  interacting  time  dependent N-electron  system
having the same density $\rho ({\bf  r},t)$. In this formalism the new
Hamiltonian, also known as the  ``Kohn-Sham" (KS) Hamiltonian, can  be
written as: 
\begin{equation}H_{KS}=\sum_i h_{KS}({\bf r}_i,{\bf p}_i,t) 
 \end{equation} where 
\begin{equation} h_{KS}({\bf r}_i,{\bf p}_i,t)=\frac{p_i^2}{2m}+
v_{ext}({\bf r}_i,t)+v_{H}({\bf r}_i,t) +v_{xc}([\rho({\bf r},t)];{\bf
r}_i,t) 
\label{veff}\end{equation} is the effective one-particle Hamiltonian.
 Apart from the external    ($v_{ext}({\bf r}_i,t)$) and  the  Hartree
($v_{H}({\bf  r}_i,t)=\int  d{\bf r'}\rho({\bf  r'},t)/|{\bf r}_i-{\bf
r'}|$)  part, the  potential contains an  ``exchange-correlation" (xc)
term  ($v_{xc}([\rho({\bf   r},t)];{\bf r}_i,t)$)   that  is an   {\it
unknown} functional of the density.    In the TDFT formalism the  wave
function   of   the  effective   noninteracting system   is  a  Slater
determinant of $N$ one-particle  orbitals $\varphi_i({\bf r},t)$ which
satisfies  the   equation:  \begin{equation}   h_{KS}({\bf     r},{\bf
p},t)\varphi_i({\bf               r},t)=i\hbar\frac{\partial}{\partial
t}\varphi_i({\bf r},t) \label{KS} .\end{equation} The particle density
can then be written as: 
\begin{equation}\rho ({\bf r},t)=\sum_i |\varphi_i({\bf r},t)|^2
.\end{equation} 

As in the time independent DFT, the main problem in TDFT  is to find a
good approximation for $v_{xc}([\rho({\bf  r},t)];{\bf  r},t)$.  Among
the most used approximations   we mention the Adiabatic Local  Density
Approximation (ALDA) \cite{ZS1},  which  is a direct extension  of the
static LDA to the time dependent problem,  and the Optimized Effective
Potential approximation (OEP)  \cite{OEP} in  which $v_{xc}([\rho({\bf
r},t)];{\bf r},t)$ is  written as a  functional of the single-particle
orbitals and (usually)  only the  exchange  part is  considered.  Both
approximations determine $v_{xc}([\rho({\bf r},t)];{\bf r},t)$ at time
$t$ as a function of the density (or  single particle orbitals) at the
{\it  same} time.   Attempts   to include the  ``memory"   of  the  xc
potential, i.e., its dependence on the  density at earlier times, have
been hampered by the fact that such a retarded potential is a severely
nonlocal functional   of the  density,   i.e., it does not  possess  a
gradient expansion  in   terms of the  density   \cite {VK1,VK2}.  For
example  an  early  attempt  by   Gross  and  Kohn  (GK) \cite{GK}  to
incorporate retardation within the  frame of the LDA  was found  to be
plagued  by inconsistencies,   such  as the   failure to  satisfy  the
``harmonic potential theorem"   \cite {D} and other  exact  symmetries
\cite {V95,VK2}.  Only very  recently \cite{VK1,VK2,VUC}  a consistent
local approximation including retardation  has been formulated  within
the  frame of the current-density functional   theory (CDFT), in which
the current-density,  rather than  the  density, is used as  the basic
variable. 

In practice, it  is not  always easy to   decide  which of the   above
approximations works  best  in a  concrete  application. A comparative
study of the  performance of different approximations  in a simple and
well controlled situation would  be very useful.  As  a first  step in
this direction, we present, in this paper,  an {\it exact} calculation
of the xc potential for what is probably the simplest nontrivial model
of interacting electrons in a  time-dependent external potential. This
model consists of  two electrons, in   two dimensions, subjected to  a
parabolic potential,  whose  curvature (which is  always  positive) is
periodically modulated  in time.  A  concrete realization of the model
could be   two electrons   in  a  quantum  dot \cite{Ash,Wag}
  with a  time-dependent
parabolic  confinement potential. We    shall show that  (i)  the time
dependent Schr\"odinger equation for  this system is  exactly solvable
by  a combination  of numerical  and  analytical methods  and (ii) the
knowledge of the exact  solution can be used  to compute the exact  xc
potential.   Our solution  for  $v_{xc}([\rho({\bf  r},t)];{\bf r},t)$
turns  out  to  be   the   time-dependent generalization of    similar
calculations recently performed  in  the static case \cite  {Fil}. The
value of these  results lies in the fact  that they provide a rigorous
benchmark, against which the merits or demerits of various approximate
theories can be assessed. 

This paper is organized as follows. In Section II we discuss the model
and   the    exact   solution  of   the   corresponding time-dependent
Schr\"odinger equation. In Section III we construct the {\it exact} xc
field   ${\bf  E}_{xc}([\rho({\bf   r},t)];{\bf  r},t)\equiv   -\nabla
v_{xc}([\rho({\bf r},t)];{\bf  r},t)$ both   in  the TDFT and  in  the
time-dependent CDFT, and we   discuss the difference between  the  two
forms.   We also   compare  our result  with  the  known static  limit
\cite{Fil}.   In Section  IV, we  draw  a comparison  between our {\it
exact} results and the ALDA, OEP, GK approximations as well as the new
approximation presented  in ref. \cite{VUC}  (VUC).   In Section V  we
introduce  a separation between  the  adiabatic and the  truly dynamic
part of  ${\bf E}_{xc}$. We conclude with   discussion and  summary
 in Section VI. 

\section{The model}

We consider two interacting electrons of effective mass $m^*$ in a 2-D
harmonic potential with  frequency  $\omega(t)$ periodic in  time. The
background dielectric constant is  $\epsilon$.  The corresponding time
dependent   Schr\"odinger     equation      in    atomic         units
($\hbar=e/\sqrt{\epsilon}=m^*=1$) is: 
\begin{equation} [-\frac{1}{2}(\nabla_1^2+\nabla_2^2) + 
\frac{1}{2}\omega^2(t)        (r_1^2+r_2^2)+\frac{1}{r_{12}}]\Psi({\bf
r_1},{\bf r_2})= i\frac{\partial}{\partial t}\Psi({\bf r_1},{\bf r_2})
\label{gen_eq} 
\end{equation} where  ${\bf r_1}$ and ${\bf  r_2}$ are  the electronic
coordinates    and $r_{12}$ is  the   distance  between the electrons.
Introducing   the Center  of   Mass  (CM)  and   Relative Motion  (RM)
coordinates ${\bf \cal R}= ({\bf r_1}+{\bf  r_2})/2$ and ${\bf r}={\bf
r_1}-{\bf r_2}$ the eq. (\ref{gen_eq}) decouples in the two equations: 
\begin{eqnarray} \label{CM1} (-\frac{1}{4}\nabla_{\bf {\cal
R}}^2+\omega^2(t){\cal                        R}^2)\Psi_{CM}({\bf{\cal
R}},t)&=&i\frac{\partial}{\partial  t}\Psi_{CM}({\bf\cal     R},t)  \\
(-\nabla_{\bf
r}^2+\frac{1}{4}\omega^2(t)r^2+\frac{1}{r})\Psi_{RM}({\bf
r},t)&=&i\frac{\partial}{\partial t}\Psi_{RM}({\bf r},t)\label{RM1} 
\end{eqnarray} where \begin{equation}\label{orb} \Psi({\bf r_1},{\bf
r_2})=\Psi_{CM}({\bf{\cal R}},t)\Psi_{RM}({\bf  r},t)\end{equation} is
the orbital part of the wave function. The spin state  can be either a
singlet or a triplet (we assume 3D isotropy for  the spin $S$). The RM
wave function must be even  or odd under  inversion ${\bf r} \to -{\bf
r}$, depending on whether $S=0$ or $S=1$ respectively. 

For simplicity of notation, we are using ``${\bf r}$'' to indicate the
RM coordinate going back to  ``${\bf r}_{12}$''  only where needed  to
avoid confusion. 
 
\subsection{Solution in the CM channel\label{a}}

\begin{description}
\item[(i)] {\bf General analytical solution } 
\end{description}
The  problem of a  quantum  harmonic oscillator  with a time dependent
frequency has been  studied by several  authors \cite{TDHO}.  Equation
({\ref{CM1}) is analytically solvable for  a general (periodic or not)
$\omega(t)$.  The  angular momentum is a  constant  of motion and this
allows the separation of angular and  radial coordinates. So we obtain
the radial equation: 
\begin{equation}\label{CMr2}
(-\frac{1}{4}\frac{\partial^2}{\partial                          {\cal
R}^2}-\frac{1}{4}\frac{1}{\cal          R}\frac{\partial}{\partial\cal
R}+\omega^2(t)           {\cal        R}^2+\frac{1}{4}\frac{m^2}{{\cal
R}^2})\chi_{n,m}({\cal  R},t)=i\frac{\partial}{\partial  t}\chi_{n,m}(
{\cal R},t).\end{equation} where 
\begin{equation}
\Psi_{CM}({\bf            {\cal                R}},t)=\chi_{n,m}({\cal
R},t)\Theta_m(\vartheta)\label{separ}\end{equation} with 
\begin{equation}\Theta_m(\vartheta)=\frac{1}{\sqrt{2\pi}}
e^{-im\vartheta},\label{teta} 
\end{equation} where $m$ is a positive integer denoting the (constant)
angular momentum and $\vartheta$  is   the angular coordinate  of  the
center of mass. 

The general solution of eq. (\ref{CMr2}) is given by: 
\begin{eqnarray} \chi_{n,m}( {\cal R},t)&=&\sqrt{\frac{n!}{2^m(n+m)!}}
\Big(\frac{d\phi}{dt}                                            \Big)
^{\frac{m+1}{2}}\exp\Big(i(2n+m+1)(\phi(0)-\phi(t))\Big)\cdot
\nonumber\\           &             &            2^m{\cal         R}^m
\exp\Big((-\frac{d\phi}{dt}+i\frac{d\ln|X|}{dt}){\cal
R}^2\Big)L_n^m\Big(\frac{d\phi}{dt}2{\cal R}^2 \Big) 
\label{chi} \end{eqnarray}
where $X(t)$ is a {\it complex} solution  of the classical equation of
motion 

\begin{eqnarray} \ddot X (t)&=&-\omega^2(t) X(t) \label{cl_eq} , 
\\ X(t)&=&|X(t)|e^{i\phi(t)} 
\end{eqnarray}
with a phase $\phi(t)$ satisfying the condition 
\begin{equation}\frac{d\phi}{dt}>0   \label{cond}.\end{equation}   The
details of  the  derivation of (\ref{chi}) are   given  in appendix A,
where it is also shown that such a solution  can always be constructed
starting   from   two    linearly  independent   real    solutions  of
eq. (\ref{cl_eq}).

We stress that eq.(\ref{teta}) and (\ref{chi})  provide a complete set
of solutions of eq.(\ref{CM1}) for whatever $\omega(t)$, provided that
the condition (\ref{cond}) is satisfied. 

In the special case  of an initial  value problem,  i.e., if  the wave
function is specified at $t=0$ as 
\begin{equation}\Psi_{CM}({\bf{\cal
R}},0)=\sum_{n,m}c_{n,m}\chi_{n,m}(                              {\cal
R},0)\Theta_m(\vartheta)\end{equation}     with  $\chi_{n,m}(    {\cal
R},0)\Theta_m(\vartheta)$ the eigenfunctions of $H(0)$, the subsequent
time evolution is given by 
\begin{equation}\Psi_{CM}({\bf{\cal
R}},t)=\sum_{n,m}c_{n,m}\chi_{n,m}(                              {\cal
R},t)\Theta_m(\vartheta)\end{equation} with  the initial condition for
$X(t)$ 
\begin{eqnarray}X(0)=\frac{1}{\sqrt{\omega (0)}} 
\\ \dot{X}(0)=i\sqrt{\omega  (0)}\end{eqnarray}  where $\omega (0)$ is
the frequency of the harmonic oscillator at the initial time.

\begin{description}
\item[(ii)]{\bf Floquet ansatz} 
\end{description}

If the Hamiltonian is periodic in time (in this case if $\omega (t+T)=
\omega (t)$, where $T$ is the period), we can look for  a basis set of
solutions satisfying the Floquet ansatz \cite{Flo1,Flo2,Flo3}: 
\begin{equation}\label{Fl_def2}\Psi(t+T)=e^{-i\varepsilon T}\Psi(t)
,\end{equation} where $\varepsilon$ (real for bound states) are called
{\it Quasi-energies} (QE).  The QE are defined modulo $\Omega=2\pi/T$.
This particular basis set has properties  that are similar to those of
the eigenstates of a static Hamiltonian \cite{Flo2}. 

In our calculations, we have chosen for $\omega^2(t)$ the form: 
\begin{equation}\omega^2(t)= \omega^2_0(1+\lambda\cos(\Omega t))  
\label{om(t)}.
\end{equation}

To construct Floquet type solutions of eq.(\ref{CMr2}) let us first of
all define the Floquet  solutions of the  classical equation of motion
(\ref{cl_eq}) \cite{Arn,Landau} as the solutions $X_{F}(t)$ having the
property 
\begin{equation}X_{F}(t+T)=e^{ iK}X_{F}(t) .\end{equation}
There exist two solutions of this kind \cite{Arn,Landau} corresponding
to  two  eigenvalues $e^{iK_{1,2}}$ (with   $K_1=-K_2$) either complex
conjugate and lying  on the unit circle  of the complex  plane or real
and  inverse to  each   other.  In   the  former  case   the solutions
$X_{F}(t)$ remain bounded in time; in the latter,
 one of them increases
exponentially for $t\to\infty$,  a phenomenon known as {\it parametric
resonance}. The actual value  of $K$ as  a function of $\lambda$  
and $\Omega$
can be calculated from eq.~ (\ref{cl_eq}). 
In this  way the $\lambda  ,\Omega$ plane can be separated
in classically stable and unstable regions.  The border of the regions
of parametric resonance are then defined by the condition: $e^{iK}=\pm
1$.

It is evident, from the form of the general solution (\ref{chi}), that
Floquet solutions  with real QE exist  within the regions of classical
stability.  In these regions the  two classical Floquet solutions  are
complex conjugate.  This means that one of the two will always satisfy
the   condition (\ref{cond}) (see   also  the equivalent  condition in
appendix A).  If we choose this  particular  solution as  the one that
determines the  time   dependence of  $\chi_{n,m}(   {\cal  R},t)$  in
eq.(\ref{chi}), then the $\{\chi_{n,m}( {\cal R},t)\}$ form a basis of
Floquet wavefunctions   with    QE: \begin{equation}\varepsilon _{n,m}
=\frac{K}{T}(2n+m+1)\label{var_CM}.\end{equation}      To       derive
eq.(\ref{var_CM})  we   made use of  the   relation : \begin{equation}
iK=\int_0^T\frac{\dot{X}}{X}dt=\int_0^T\frac{d}{dt}\ln|X|dt+i\int_0^T
\frac{d\phi}{dt}dt=i(\phi(T)-\phi(0)).\end{equation} The last equality
holds because $\int_0^Td\ln|X|/dt\cdot   dt=0$  since  $\ln|X(t)|$  is
periodic.   At the  border   of the  classical regions  of instability
$e^{iK}=\pm 1$,  the Floquet solutions $\chi _{n,m}(R,t)\rightarrow 0$
and the  set of QE $\{\varepsilon _{n,m}\}$  degenerates to $\{0\}$ if
$e^{iK}=1$ or to $\{0,\pi/T\}$ if $e^{iK}=-1$. 

In  the remaining regions   of classical instability solutions of  the
Floquet type cannot be constructed.  These regions are centered around
the values  $\Omega=2\omega_0/k$,  $k=0,1,2...$  (with $\omega_0$  the
frequency of  the unperturbed  harmonic oscillator) \cite{Arn,Landau},
so  that  the value  $\Omega=0$   is  an  accumulation  point for  the
sequence.  This  means that if   $\lambda>0$, it  is not  possible  to
perform the limit  $\Omega\to0$ without entering regions of parametric
resonance  and so it  is  not possible to follow  the  evolution of  a
Floquet state from  a finite $\Omega$  down to 0.   We remark that the
occurrence  of classical parametric resonance  is related to a failure
of the conventional Floquet theorem, which  ensures the existence of a
complete set  of  Floquet states.   The  reason is  that  our harmonic
oscillator  potential, being  not bounded,  gives  rise to  a strictly
hermitian Hamiltonian and allows   only Floquet states with real   QE:
evidently, such quasi-periodic states cannot describe the motion of an
electron  to larger  and larger  distance  from  the center that   the
resonance process   would imply.   Realistic  bounded potentials avoid
this  problem by allowing  the possibility of complex  QE in which the
electron can escape to infinity (ionization). 

\subsection{Solution in the RM channel}

As we  did  for the  CM channel,  we  separate the angular  and radial
coordinates in eq. (\ref{RM1}) and we obtain the radial equation: 
\begin{equation}\label{RMr}
(-\frac{\partial^2}{\partial  r^2}-\frac{1}{r}\frac{\partial}{\partial
r}+\frac{1}{4}\omega^2(t)  r^2+\frac{1}{r}+\frac{l^2}{r^2})\psi_{n,l}(
r,t)=i\frac{\partial}{\partial t}\psi_{n,l}( r,t).\end{equation} where 
\begin{equation}
\Psi_{RM}({\bf r},t)=\psi_{n,l}( r,t)\Theta_l(\vartheta)\end{equation}
with 
\begin{equation}\Theta_l(\vartheta)=\frac{1}{\sqrt{2\pi}}
e^{-il\vartheta} 
\end{equation} and $l$ a positive integer, even for $S=0$ and odd 
for $S=1$.   Here $\vartheta$ is  the angular  coordinate for  the  RM
channel. Eq.  (\ref{RMr}) cannot be  solved analytically except in the
following special cases: 
\begin{enumerate}
\item {\it Time independent  case} The static limit ($\lambda=0$)  has
been well analyzed in 3-dimension \cite{Fil} and for certain values of
the frequency $\omega_0$ of  the unperturbed harmonic oscillator it is
possible  to  have a completely   analytical solution also  for the RM
channel (see  \cite{Taut}). Similarly it is  possible  to construct an
analytical solution in the 2-dimensional case (see appendix B). 
\item {\it Weak correlation  limit}   We define the  weak  correlation
limit as the regime   in which the  Coulomb interaction  is negligible
compared to the harmonic confinement potential. This means that 
\begin{equation}
 \frac{l}{a}\to 0 
\end{equation}
where  $l\equiv  \sqrt{\hbar/2m^*\omega_0}$ is the  confinement length
due to the harmonic  potential and $a\equiv \hbar^2\epsilon/m^*e^2$ is
the        effective     Bohr    radius.          In      our    units
($\hbar=e/\sqrt{\epsilon}=m^*=1$)  this   is  equivalent  to  imposing
$\omega_0\to\infty$.   In   this regime   the  coulombic term  becomes
negligible and the RM problem  becomes analytically solvable (see part
A of this section). 
\item     {\it Strong correlation   limit     in the linear   response
approximation  with respect  to  $\lambda$} In this  limit the Coulomb
interaction dominates the harmonic confinement potential. This means 
\begin{equation}\frac{l}{a}\to \infty \label{stro}\end{equation}
(so in our units  $\omega_0\to 0$) and  the two electrons can be shown
to perform small oscillations about the classical equilibrium position
determined  by  the  competition between  electrostatic  repulsion and
harmonic confinement. Expanding  the potential  energy to the   second
order in the displacement from the classical equilibrium distance $r_0
\gg  l$ and  neglecting  corrections of  order $l/r_0$ to  the kinetic
energy one obtains the effective harmonic Hamiltonian 
\begin{equation} H_{eff}(t)=\frac{p^2_r}{2\mu}+\frac{\mu}{2}
\tilde{\omega}^2(t)(r-r_0)^2-\mu E_{ext,1}(t)(r-r_0)\label{approRM1} 
\end{equation} 
where  $ p^2_r=-\partial^2/\partial r^2$,   $\mu=1/2 $ is  the reduced
       mass,          $\tilde{\omega}^2(t)=3\omega_0^2+\omega_1^2(t)$,
       $\omega_1^2(t)$ is defined as 
\begin{equation}\omega^2_1(t)\equiv \omega^2_0\lambda\cos(\Omega t),
\label{om_1}\end{equation} and $E_{ext,1}(t)=-\omega_1^2(t)r_0$ 
can be  viewed as  an ``external force''.  Apart  from time  dependent
phase  factors  (see Appendix C  for these factors and for details   
of  the derivation) the
solution, for $n=l=0$, takes the form: 
\begin{equation} \Psi(r,t)=\frac{1}{\pi^{\frac{1}{4}}}
(\frac{d{\tilde\phi}}{dt})^{\frac{1}{4}}
e^{-i\mu\dot{x}_0(t)(r-r_0)}e^{\frac{i}{2}\mu       \frac{\dot{{\tilde
X}}}{{\tilde X}}(r-r_0+x_0(t))^2} 
\label{highwf}
\end{equation}
with ${\tilde X}(t)$ the solution of 
\begin{equation}\ddot{{\tilde X}}=-\tilde{\omega}^2(t){\tilde X}(t)
\label{tilde X},\end{equation} ${\tilde \phi}(t)$ its phase such that 
$d{\tilde\phi}/dt>0$ and $x_0(t)$ the solution of 
\begin{equation}\ddot{x}_0=-E_{ext,1}(t)-\tilde{\omega}^2(t)x_0(t)
\label{x_0}.
\end{equation}
 If we  insert in eq.  (\ref{highwf}) for $x_0(t)$ its linear response
 approximation expression 
\begin{equation}  x_0(t)=-\frac{E_{ext,1}(t)}{3\omega_0^2-\Omega^2}
\label{x_0lin}\end{equation}
and for ${\tilde X}(t)$  the classical Floquet solution of (\ref{tilde
X}) with $d{\tilde\phi}/dt>0$, we obtain a Floquet solution for the RM
problem. 
\end{enumerate}
In the general  case eq. (\ref{RMr})  must be solved numerically, and,
to construct ${\bf E}_{xc}({\bf r},t)$, the  most natural choice is to
consider the dynamical  equivalent of the ground  state, that  is, for
the     RM      channel,    the      ``lowest''      Floquet     state
$\Psi_{RM}^0(r,t)=\psi_{0,0}( r,t)/\sqrt{2\pi}$. We define this as the
state that  evolves continuously from the  ground state of  the static
Hamiltonian    as  the amplitude   $\lambda$   of  the time  dependent
perturbation grows from zero. From now  on we only consider $l=0$ (and
correspondingly $m=0$ for the CM channel). 

In order to calculate this Floquet state we  use its property of being
an eigenstate  of the one-period  time-evolution  operator $\hat U(T)$
($\Psi({\bf r},t+T)=\hat    U(T)\Psi({\bf  r},t)$)   with   eigenvalue
$e^{-i\varepsilon  T}$   (see eq.  (\ref{Fl_def2})).   The  idea is to
calculate the matrix $\{U(T)_{ij}\}$ in  a suitable basis, diagonalize
it and find its ``lowest''  eigenstate - the ``lowest'' Floquet state.
The  basis we  choose  to calculate  $\{U(T)_{ij}\}$  is   the set  of
eigenstates  of  a  two dimensional  harmonic  oscillator with angular
momentum equal  to  zero $\{R_i(r,0)\}$.  For  a  general instant $t$,
$\{U(t)_{ij}\}$ are defined by the equation: 
\begin{equation} \label{exp} R_j(r,t)=\sum_{i=1}^MU(t)_{ij}R_i(r,0),
\end{equation}
where  the  sum has  been  truncated for  practical purposes.   In our
 calculation $M=60$ - a value that ensures  a very good convergence of
 the lowest QE's.   Inserting   for each  $R_j(r,t)$  the   expression
 (\ref{exp}) into eq. (\ref{RMr}), we find for $U(t)_{ij}$ a system of
 $M$ first order differential equations.  Integrating this system over
 one period, {\it  for  each} $R_j(r,t)$, we  obtain  $\{U(T)_{ij}\}$.
 Since the QE $\varepsilon_j$ are defined modulo $\Omega$ \cite{Flo3},
 it is not possible to establish from the  value of the eigenvalues of
 $U(T)$ the  ``lowest'' one. To identify  it we  have instead used the
 property that  for $\lambda\to 0$, $\varepsilon_j\to\varepsilon_j^0$,
 where $\varepsilon_j^0$ is an eigenvalue of the static ($\lambda= 0$)
 Hamiltonian  \cite{Flo2,Flo4}.   In  practice  we  have  followed the
 evolution of the ground state energy $\varepsilon_0^0$ for increasing
 $\lambda$'s.

\section{Construction and calculation of the exact 
exchange-correlation potential} {\bf Construction in TDFT} 

If we consider  two electrons  in a  singlet  state, the KS  equations
reduce to     a single  equation for     the doubly  occupied  orbital
$\varphi({\bf r},t)$ 
\begin{equation}(-\frac{1}{2}\nabla^2+v_{ext}({\bf r},t)
+v_{H}({\bf  r},t)+v_{xc}([\rho({\bf   r},t)];{\bf r},t)) \varphi({\bf
r},t)=i\frac{\partial}{\partial t}\varphi({\bf r},t) 
\label{KSphi}.\end{equation}
The KS orbital can be written as 
\begin{equation}\varphi({\bf r},t)=|\varphi({\bf r},t)|
e^{if({\bf r},t)}
\label{phiorb}\end{equation}
and its modulus is related to the density by the equation: 
\begin{equation} \rho({\bf r},t)=2|\varphi({\bf r},t)|^2 \label{rho}
\end{equation} 
while   its phase is  related to  the  KS  velocity ${\bf v}_{KS}({\bf
r},t)$ by 
\begin{equation}\nabla f({\bf r},t)\equiv {\bf v}_{KS}({\bf r},t).
\end{equation}
If we  insert  expression  (\ref{phiorb}) in  eq.(\ref{KSphi})  and we
impose that $v_{xc}([\rho({\bf  r},t)];{\bf r},t)$ is  real, we obtain
two equations, one from the real part of (\ref{KSphi}): 
\begin{eqnarray}& & \frac{1}{4}\nabla^2\ln \rho({\bf r},t)+\frac{1}{8}
|\nabla\ln   \rho({\bf r},t)|^2-v_{ext}({\bf  r},t)-v_H({\bf  r},t)  -
v_{xc}([\rho({\bf r},t)];{\bf            r},t)\nonumber  \\   &      &
-\frac{1}{2}|\nabla  f({\bf r},t)|^2-\frac{\partial}{\partial t}f({\bf
r},t)\label{vxcD}=0 \end{eqnarray}  and  the second from its imaginary
part 
\begin{equation} \nabla\cdot\nabla f({\bf r},t)+\nabla f({\bf r},t)
\cdot\nabla\ln  \rho({\bf     r},t)+\frac{\partial}{\partial    t} \ln
\rho({\bf r},t)=0  .\label{pdef}\end{equation}  Eq.(\ref{vxcD}) can be
solved for $v_{xc}({\bf  r},t)$  (for simplicity  of  notation we have
dropped the dependence of the xc potential on the density) and we find
the  following  explicit expression for   the xc  electric field ${\bf
E}_{xc}({\bf r},t) \equiv -\nabla v_{xc}({\bf r},t)$: 
\begin{eqnarray}{\bf E} _{xc}({\bf r},t)&=&-\nabla(\frac{1}{4}
\nabla^2\ln \rho({\bf r},t)+\frac{1}{8}|\nabla\ln \rho({\bf  r},t)|^2)
-{\bf  E}_{ext}({\bf r},t)-{\bf  E}_H({\bf  r},t)  \nonumber  \\ &   &
-\nabla(-\frac{1}{2}|\nabla  f({\bf  r},t)|^2-\frac{\partial}{\partial
t}f({\bf r},t))\label{excD},\end{eqnarray} where  ${\bf E}_{H}=-\nabla
v_{H}$  and ${\bf E}_{ext}=-\nabla v_{ext}$.   The  last two terms  of
eq.(\ref{excD})  are peculiar of the  time dependent problem while the
first  three correspond to the  static expression \cite{Fil} for ${\bf
E}_{xc}$, 
\begin{equation}{\bf E} _{xc}^{static}({\bf r})=
-\nabla(\frac{1}{4}\nabla^2\ln  \rho({\bf  r})+\frac{1}{8}  |\nabla\ln
\rho({\bf r})|^2)-{\bf E}_{ext}({\bf r})-{\bf E}_H({\bf r}) 
\label{statexc}.\end{equation}
Eq.(\ref{pdef})  is a  first order  partial  differential equation for
$\nabla f$ and is  equivalent to the  continuity equation for the {\it
non} interacting   KS  system.    It  shows that   ${\bf  v}_{KS}({\bf
r},t)=\nabla f({\bf   r},t)$   is  in  general a    {\it  non} trivial
functional of the density. We stress that ${\bf v}_{KS}({\bf r},t)$ is
in general not the  same as the  {\it exact} velocity field;  only the
longitudinal part  of   the KS   current    must  coincide with    the
longitudinal  part   of the  physical  current   due to the continuity
equation.

{\bf Construction in time dependent CDFT} 

The time dependent CDFT  differs from  the  TDFT in that not  only the
density  but also the  current density calculated  from  the KS single
particle orbitals is exact. 

In  order to accomplish this, one  introduces an  xc vector potential,
${\bf A}_{xc}$  in   the Kohn-Sham equation  \cite{VK1,VK2}.    The KS
Hamiltonian \cite{VK2} is now: 
\begin{equation}H^{KS}(t)=\sum_i (\frac{1}{2}({\bf p}_i+{\bf
A}_{xc}({\bf           r_i},t))^2+v_{H}({\bf      r_i},t)+v_{ext}({\bf
r_i},t))\end{equation} which yields {\it both} the correct density and
current. In the case of two electrons in a singlet  state, we get, for
the occupied  orbital   $\varphi({\bf  r},t)$,   the  time   dependent
Schr\"odinger equation: 
\begin{equation}((\frac{1}{2}({\bf p}+{\bf
A}_{xc}({\bf           r},t))^2+v_{H}({\bf          r},t)+v_{ext}({\bf
r},t)))\varphi({\bf   r},t)=i\frac{\partial}{\partial   t}\varphi({\bf
r},t)\label{SchCDFT}\end{equation} and  we   can now follow  the  same
procedure used  in  TDFT to  find   an explicit  expression for  ${\bf
E}_{xc}({\bf r},t)$. We obtain: 
\begin{eqnarray} {\bf E}_{xc}({\bf r},t)&=&-{\bf 
\dot{A}}_{xc}({\bf         r},t)              \nonumber             \\
&=&-\nabla(\frac{1}{4}\nabla^2\ln                            \rho({\bf
r},t)+\frac{1}{8}|\nabla\ln \rho({\bf     r},t)|^2)-{\bf E}_{ext}({\bf
r},t)-{\bf      E}_H({\bf      r},t)      \nonumber     \\      &    &
+\nabla(\frac{1}{2}v^2)+{\bf \dot{v}}\label{excC},\end{eqnarray} where
${\bf  v}({\bf r},t)$ is the  exact  velocity of the {\it interacting}
system, ${\bf v}({\bf r},t)=\nabla f({\bf r},t) + {\bf A}({\bf r},t)$.
From the imaginary part  of eq. (\ref{SchCDFT}) (or equivalently  from
the  continuity equation) we get  the first order partial differential
equation 
\begin{equation} \nabla\cdot {\bf v}({\bf r},t)+{\bf v}({\bf r},t)
\cdot\nabla\ln   \rho({\bf    r},t)+\frac{\partial}{\partial   t}  \ln
\rho({\bf r},t)=0.\end{equation} 

The  advantage   of this   formulation  is  that   it expresses  ${\bf
E}_{xc}({\bf r},t)$ as  a function  of  the physical  quantities ${\bf
v}({\bf r},t)$ and $\rho({\bf r},t)$. 

{\bf Circularly symmetric states}

If the time dependent state is circularly symmetric, as in the case we
 are studying,  then  the current  is purely radial,  therefore purely
 longitudinal, and the two  expressions (\ref{excD})  and (\ref{excC})
 coincide (that  is  ${\bf  v}({\bf   r},t)  \equiv{\bf   v}_{KS}({\bf
 r},t)$). Thus, in this case, there  is no difference between the time
 dependent DFT and  CDFT.   Eq. (\ref{pdef}) can be  easily integrated
 yielding 
\begin{equation}\frac{\partial}{\partial r}f(r,t)=
-\frac{1}{r\rho(r,t)}
\int_0^r\frac{\partial \rho(r',t)} {\partial t}r'dr'.\end{equation}

{\bf Linear response} 

In    the limit     of   small external  time-dependent   perturbation
($\lambda\to 0$ in eq.  (\ref{om(t)})),  we expand all the  quantities
to  first order  in $\lambda$,  i.e.  ${\bf E}_{ext}={\bf E}_{ext ,0}+
{\bf  E}_{ext ,1}$, ${\bf E}_H={\bf  E}_{H  ,0}+ {\bf E}_{H,1}$, ${\bf
E}_{xc}={\bf E}_{xc ,0}+  {\bf E}_{xc ,1}$  and $\ln \rho =\ln  \rho_0
+\rho_1/\rho_0$   where  the   subscripts  ``0'' and    ``1'' indicate
respectively zero and first order with respect to $\lambda$. Then from
eq.   (\ref{excD})   we  obtain  \begin{equation}    {\bf  E}_{xc,0}(r
)=-\nabla(\frac{1}{4}\nabla\ln            \rho_0+\frac{1}{8}|\nabla\ln
\rho_0|^2)-{\bf                                     E}_{ext,0}(r)-{\bf
E}_{H,0}(r)\label{xc,0}\end{equation} and 
\begin{equation} {\bf E}_{xc,1}(r,t)=\nabla \frac{\partial}
{\partial     t}f(r,t)-\frac{1}{4}\nabla(\nabla^2\frac{\rho_1}{\rho_0}
+\nabla\ln   \rho_0\cdot\nabla\frac{\rho_1}{\rho_0})-{\bf  E}_{ext,1}(
r,t)  -{\bf  E}_{H,1}(r,t)\label{xc,1},\end{equation}  where   we have
neglected also the terms of order $v^2(r,t)=|\nabla f(r,t)|^2$. 
 
{\bf Calculation of the exact ${\bf E}_{xc}(r,t)$}  

We  have  considered in  detail  two  sets  of the parameters  
$\omega_0$,
$\Omega$ and $\lambda$  that appear in eq.(\ref{om(t)}),  
corresponding to  high
and   low  correlation.   The     two   sets  are  given    in   Table
\ref{table1}. The values of $\omega_0$  have been chosen such that it
is  possible  to   construct   analytically  the   solution   of   the
corresponding static   Schr\"odinger equation  (see  appendix  B). The
values of  $\lambda$ and $\Omega$ have  been chosen so that the system
is in  the   linear response regime  but  well  above  the  regions of
parametric resonance for the   CM channel and above  
the first excitation energy of the system.   

The exact time dependent densities are plotted in Fig.1 and Fig.2. For
the weak correlation parameter $\omega_0=1$ the density is centered at
the origin   as we can  expect from  the  exact solution  in  the weak
correlation   limit (a Gaussian  centered  at  the origin, see section
II). In the case of high correlation  ($\omega_0\approx 0.02$), on the
other hand,  the maximum of  the density  is  at {\it finite} distance
from  $r=0$  in agreement  with  the form  that  the  RM wave function
assumes in the  strong correlation limit eq.(\ref{highwf}) (an annulus
of average   radius  $r_0/2$, with  $r_0$   the  classical equilibrium
distance of the two electrons): the increased  strength of the Coulomb
repulsion   in respect to   the   harmonic confinement pushes on 
average the two
electrons   far from each other.   In these  plots the solid
line represents   the static  limit while  each   of the broken  lines
corresponds to the time dependent $\rho(r,t)$ at different times. 

In the insets of Fig.1  and Fig.2 we  plot the time dependent velocity
${\bf v}(r,t)=\nabla f(r,t)$, that is  the other necessary  ingredient
to  calculate ${\bf E}_{xc}(r,t)$.  As the   plots show, the motion is
approximately a  ``breathing''  motion: the  velocity  is zero at  the
origin   while for  $r\neq 0$  it   increases  almost linearly.    The
asymptotic behavior is linear in $r$ with a correction in $1/r^2$.  In
Fig.3 ($\omega_0=1$) and   Fig.4 ($\omega_0\approx 0.02$)   we finally
plot  the  results for the field  ${\bf  E}_{xc}(r,t)$. The solid line
represents the static limit while each of the broken lines corresponds
to  ${\bf  E}_{xc}(r,t)$ at different times.   We  choose to plot this
quantity instead of  the  more traditional $v_{xc}({\bf r},t)$   since
this is  the  meaningful physical  quantity whose asymptotic  behavior
does not depend on arbitrarily fixed time dependent constants.  As can
be seen  from the plots, as  the correlation  in the system increases,
the positive peak   of  ${\bf E}_{xc}(r,t)$  for small   $r$ increases
too. This is related to the  enhancement
 of the  strength of the Coulomb
repulsion  that, as seen 
before,  pushes  the
maximum of the density away from the origin.
 ${\bf E}_{xc}(r,t)$ can be
viewed as a    force  in the   KS   system which, where   positive,
contributes to drag the particles away from the
origin.
 
Starting from the asymptotic form of the RM  wave function for $r_{12}
\to         \infty$,      that            is,        $\Psi_{RM}\propto
r_{12}^{\alpha}\cdot(1+(b_{\Re}+ib_{\Im})/r)\cdot
\exp(-r_{12}^2\dot{\phi}/4   )$,       with         $\alpha$     real,
$b_{\Im}=-(\dot{b}_{\Re}+b_{\Re}\ddot{\phi}/
(2\dot{\phi}))/\dot{\phi}$    and     $b_{\Re}$       defined       by
$\ddot{b}_{\Re}=-\omega^2(t)  b_{\Re}+\dot{\phi}$,   we get,   with  a
little algebra, the asymptotic form  of the density ($\rho(r,t)\propto
r^{2\alpha}(1+2b_{\Re}/r)\cdot \exp(-r^2\dot{\phi})$)  and    of   the
velocity         ($\nabla       f(r,t)\approx-r/2\cdot           d(\ln
\dot{\phi})/dt-b_{\Im}/r^2)$).  From  these   behaviors,  using    eq.
(\ref{excD}),  we can derive the  asymptotic behavior for the xc field
that  results  to   be the   same as   in   the   static case:   ${\bf
E}_{xc}\approx-1/r^2$ for $r \to \infty$. 

In Fig.3 and  Fig.4 we also plot,  for comparison,  the
Local Density Approximation (LDA) of the static ${\bf E}_{xc}(r)$.  As
can be  seen from the figures, LDA  behaves, in general, reasonably
 well
except for small $r$ (for which in the  weak correlation case has even
the wrong sign) and for large $r$ (for which decreases exponentially).

\section{Comparison with approximate theories}
In this section we discuss the comparison between our exact result for
${\bf  E}_{xc}(r,t)$  and  the  results obtained  from  the  most used
approximations,  namely  the   Adiabatic Local   Density Approximation
(ALDA), the  Optimized Effective Potential (OEP),  the Gross  and Kohn
(GK)  approximation, and   the   hydrodynamic  approximation  recently
introduced  by Vignale, Ullrich, and Conti  (VUC). The expressions for
the xc electric field in these approximations are: 
\begin{itemize}
\item ALDA \cite{ZS1}: 
\begin{equation} {\bf E}_{xc}^{ALDA}(r,t)=-\nabla(\frac{d}{d\rho}
(\rho(r,t)\varepsilon_{xc}(\rho)))\end{equation}                 where
$\varepsilon_{xc}(\rho) $ is the xc energy per particle of the 
homogeneous electron
gas. In 2D it is given by: 
\begin{eqnarray}\varepsilon_{xc}&=&-\frac{4\sqrt{2}}{3\pi r_s}+
\frac{a_0}{2}\frac{1+a_1\sqrt{r_s}}{1+a_1\sqrt{r_s}+a_2r_s+a_3r_s^
{\frac{3}{2}}}\cdot \frac{e^4m^*}{\hbar^2\epsilon},    \\    
r_s&=&\frac{1}{\sqrt{\pi\rho}},\end{eqnarray}
$a_0=-0.3568,~a_1=1.1300,~a_2=0.9052,~a_3=0.4165$\cite{Isi,TaCe} 
\item OEP approximation  (which in this simple  case is equivalent  to
the Hartree Fock approximation)\cite{OEP,GDP} 
\begin{equation} {\bf E}_{xc}^{OEP}(r,t)=-\nabla 
(-\frac{1}{2}v_H(r,t))
\end{equation}
\item   GK approximation \cite{GK}   (valid   in the linear   response
regime): 
\begin{equation} {\bf E}_{xc,1}^{GK}(r,\omega)=
-\nabla(\rho_1(r,\omega)
f_{xc}(\rho_0,\omega))\end{equation} where ${\bf  E}_{xc,1}(r,\omega)$
is   a  Fourier  component  of  ${\bf   E}_{xc,1}(r,t)$ (defined, with
$\rho_1(r,t)$,   in   the     ``Linear   response''    section)    and
$f_{xc,L}(\rho_0,\omega)$  is  the longitudinal part of  the frequency
dependent $xc$ kernel of the homogeneous electron gas. 
\item   VUC  approximation \cite{VUC}   for   a  circularly  symmetric
potential in 2D (valid in the linear response regime): 
\begin{eqnarray}{\bf E}_{xc,1}^{VUC}(r,\omega)&=&{\bf E}_{xc}^{ALDA}
(r,\omega)+  \frac{1}{\rho_0}(\nabla(\rho_0^2(f_{xc,L}(\rho_0,\omega)-
\frac{d^2\rho\varepsilon_{xc}}{d\rho^2})(\frac{v_1(r,\omega)}{r}
+\frac{\partial  v_1(r,\omega)}   {\partial    r}))\nonumber    \\&  &
-\frac{2}{r}\nabla(\rho_0^2f_{xc,T}(\rho_0,\omega))v_1(r,\omega) ) 
\label{eqVUC}\end{eqnarray}
where     $v_1(r,t)$  is    the   velocity    field  and
$f_{xc,L}(\rho_0,\omega)$,    $f_{xc,T}(\rho_0,\omega)$   are      the
longitudinal and the transverse  part of the frequency  dependent $xc$
kernel of the homogeneous electron gas. 

In   both GK  and VUC  we   have used  for $f_{xc}(\rho_0,\omega)$ the
expressions recently obtained by Nifos\'i et al.  \cite{NCT}. 
\end{itemize}
The  comparison between  the {\it  exact}  ${\bf E}_{xc}(r,t)$ and its
approximations  is made  plotting  its  first Fourier component  ${\bf
E}_{xc}(r,\Omega)$.   Since our calculations were   done in the linear
response regime, the difference  between ${\bf E}_{xc}(r,\Omega)$  and
${\bf E}_{xc,1}(r,\Omega)$ is negligible. 

In  the case of  weak  correlation (Fig.  5,  $\omega_0=1$)  the ALDA,
reproduces,   except for  small   $r$,   the  general trend,    though
underestimating  the peak of the potential.  In the strong correlation
case (Fig. 6,  $\omega_0\approx 0.02$) it underestimates the potential
for very  small $r$, but, for intermediate  values, it gets  closer to
${\bf E}_{xc}(r,\Omega)$.  For weak  correlation and for  small values
of $r$  the OEP approximation does  not  reproduce the exact behavior,
while in the region in which ${\bf E}_{xc}(r,\Omega)$ is significantly
non  zero it gets  closer to the exact  result.   In the limit of zero
correlation the OEP, which is   equivalent to Hartree-Fock theory  for
this   system, would  give the  exact   result.  Its  behavior gets
worse  when  the correlation increases  (see Fig.  6):  it is the only
approximation  that does not even  reproduces  the first peak of ${\bf
E}_{xc}(r,\Omega)$.   On  the  other  hand,   this is   the {\it only}
approximation  that has the correct  asymptotic  behavior $-1/r^2$ for
$r\to\infty$.  In   the   weak correlation  case    (Fig.  5)  the  GK
approximation  has a   behavior similar  to the  OEP   (except for the
asymptotic behavior  that is  not  reproduced  correctly), while,  for
strong   correlation (Fig. 6)  it   reproduces  the correct trend  but
underestimates ${\bf E}_{xc}(r,\Omega)$   for small values of $r$  and
overestimates  it for  intermediate   values.  In  the  case  of  weak
correlation (Fig.  5)  the VUC approximation   does not reproduce  the
exact trend  for  small $r$,   while, for  intermediate values it  get
closer to   ${\bf E}_{xc}(r,\Omega)$ though  underestimating its peak.
For strong correlation  its behavior is almost indistinguishable  from
the ALDA.

\section{``Adiabatic'' and ``dynamic'' exchange correlation potentials}

There  has been considerable   effort, in recent   years, aimed at the
construction of a fully dynamic  xc potential, which, unlike the  ALDA
potential, should depend  on the density at  all previous times, i.e.,
have  a memory.  In  order to assess  the  importance of these 
``memory
effects'', we  shall now
separate  ${\bf E}_{xc}(r,t)$  in an  ``{\it  adiabatic}'' part  ${\bf
E}_{xc}^{ad}(r,t)$  containing the  adiabatic  evolution of   the {\it
static  exact} ${\bf E}_{xc}(r)$  and a ``{\it  dynamical}'' one ${\bf
E}_{xc}^{dy}(r,t)$, peculiar of the time dependent problem.  Comparing
equations (\ref{excD}) and   (\ref{statexc}), it is easy  to  identify
$-\nabla(\nabla^2\ln    \rho(r,t)/4+|\nabla\ln    \rho(r,t)|^2/8)-{\bf
E}_H(r,t))$   as       adiabatic     terms,  while   $-\nabla(-|\nabla
f(r,t)|^2/2-\partial f( r,t)/\partial t)$  are  peculiar to  the  time
dependent  case. However  ${\bf   E}_{ext}(r,t)$ is  not  an  explicit
functional  of the  time  dependent density and  must  be treated more
carefully.  The  adiabatic part of  ${\bf E}_{ext}(r,t)$ is defined as
the electric field ${\bf  E}_{ext}^{ad}(r,t)$  which, when applied  to
the   physical interacting   system,  would yield    the exact density
$\rho(r,t)$ while the system remains in the instantaneous 
ground state.

It is then possible   to define the  ``{\it  dynamical}'' part of  the
external field as what remains after subtracting the adiabatic part: 
\begin{equation} {\bf E}_{ext}^{dy}(r,t)\equiv 
{\bf E}_{ext}(r,t)-{\bf E}_{ext}^{ad}(r;t)  .\end{equation} Now we can
separate, in the case of the two electron problem, ${\bf E}_{xc}(r,t)$
in  an    adiabatic (${\bf E}_{xc}^{ad}(r,t)$) and    dynamical (${\bf
E}_{xc}^{dy}(r,t)$)    part,    ${\bf     E}_{xc}(r,t)=           {\bf
E}_{xc}^{ad}(r,t)+{\bf E}_{xc}^{dy}(r,t)$, where 
\begin{eqnarray}{\bf E}_{xc}^{dy}(r,t)&=&-\nabla(-\frac{\partial 
f(r,t)}    {\partial    t}-\frac{1}{2}(\nabla      f(r,t))^2 )-   {\bf
E}_{ext}^{dy}(r,t) 
\label{vxc_na}\\          {\bf             E}_{xc}^{ad}(r,t)&=&-\nabla
(\frac{1}{2}[(\frac{1}{2}\nabla               \ln               (\rho(
r,t)))^2+\frac{1}{2}\nabla^2 (\ln (\rho( r,t)))]) \nonumber \\& &-{\bf
E}_H(r,t) -{\bf E}_{ext}^{ad}(r,t) 
\label{vxc_ad}.\end{eqnarray}

In the linear regime, using the  linearized expression (\ref{xc,1}) we
obtain: 
\begin{eqnarray}{\bf E}_{xc,1}^{ad}(r,t)&=&-\nabla(\frac{1}{4}
\nabla^2\frac{\rho_1}{\rho_0}+\frac{1}{4}\nabla  \ln  \rho_0    \nabla
\frac{\rho_1}{\rho_0})-{\bf E}_{ext,1}^{ad}(r,t)-{\bf E}_{H,1}(r,t) 
\label{xc,1,ad}\\
{\bf    E}_{xc,1}^{dy}(r,t)&\equiv     &{\bf       E}_{xc,1}(r,t)-{\bf
E}_{xc,1}^{ad}(r,t)\\   &=&  {\dot{\bf v}}-{\bf   E}_{ext,1}(r,t)+{\bf
E}_{ext,1}^{ad}(r,t) \label{xc,1,dy} 
\end{eqnarray}
where   we have used the fact   that $\nabla  f(r,t)={\bf v(r,t)}$ and
neglected terms of order $v^2(r,t)$. 

The difficulty in the calculation of ${\bf E}_{ext}^{ad}(r,t)$ is that
in  general its form  is unknown and   leads to a  non
separable, two-electron Schr\"odinger   equation.  In our  case  it is
possible to  calculate analytically ${\bf E}_{ext}^{ad}(r,t)$  and its
counterpart ${\bf E}_{ext}^{dy}(r,t)$ in the  limit of extremely  weak
and extremely strong correlation, but for  a general set of parameters
it  will  be   necessary  to  find  an   approximation  for ${\bf
E}_{ext}^{ad}(r,t)$. 

We   will now show that   ${\bf  E}_{xc,1}^{dy}(r,t)$, in this system,
vanishes exactly  in both the  weak and strong correlation limits, and
it is likely to be very small in the intermediate cases. 

{\bf Calculation of ${\bf E}_{xc,1}^{dy}(r,t)$} 

{\bf ``Weak'' correlation limit} 

In this regime the response of our system to the external potential 
\begin{equation}v_{ext}(r,t)=\frac{1}{2}(\omega_0^2+\omega_1^2(t))r^2 
\end{equation}
with $\omega_1^2(t)$ defined  by  eq.  (\ref{om_1}), is  a ``breathing
motion'', i.e. it can be described as a periodic transformation with a
length scale 
\begin{equation} l(t)\propto (\frac{d}{dt}\phi(t))^{-\frac{1}{2}}
\propto |X(t)|\label{breat}. \end{equation} 

Then   we  can calculate explicitly all    the quantities appearing in
eq.(\ref{xc,1,dy}).  The velocity field is given by 
\begin{eqnarray}{\bf v}(r,t)&=&\frac{\dot{l}(t)}{l(t)}{\bf \hat{r}}=
(\frac{d}{dt}|X(t)|){\bf              \hat{r}}\\             {\dot{\bf
v}(r,t)}&=&(\frac{d^2}{dt^2}\ln(|X(t)|)){\bf                \hat{r}}\\
&=&(\frac{1}{|X(t)|}\frac{d^2}{dt^2}|X(t)|-(\frac{1}{|X(t)|}
\frac{d}{dt}
|X(t)|)^2) {\bf \hat{r}},\label{acce}\end{eqnarray} the external field
is given by 
\begin{equation}
{\bf  E}_{ext,1}(r,t)=  -\omega_1^2(t){\bf r},\end{equation} and   the
``adiabatic'' external field is given by 
\begin{equation}
{\bf                 E}_{ext,1}^{ad}(r,t)=-(\frac{d}{dt}\phi(t))^2{\bf
r}\label{adas}.\end{equation}     This is  also     the  exact   ${\bf
E}_{ext,1}^{ad}(r,t)$ corresponding     to the noninteracting harmonic
oscillator problem. 

We    can   now  prove that   ${\bf     E}_{xc,1}^{dy}(r,t)=0$ in this
approximation.  Using in eq.(\ref{acce}) $|X(t)|=X(t)\exp(-i\phi(t))$,
${\ddot X}=-\omega(t)^2X$ and  dropping the terms  of second order  in
$\lambda$, we get: 
\begin{eqnarray}{\dot{\bf v}(r,t)}&=&(-\omega_1(t)^2+\frac{d}{dt}
\phi(t)^2){\bf           r}\\     &=&{\bf         E}_{ext,1}(r,t)-{\bf
E}_{ext,1}^{ad}(r,t)\end{eqnarray}       that      substituted      in
eq.(\ref{xc,1,dy}) yields ${\bf E}_{xc,1}^{dy}(r,t)=0$. 

{\bf ``Strong'' correlation limit} The  equilibrium density reduces to
 a $\delta$-shell $\rho_0=2/(\pi r_0)\delta(r-r_0/2)$ and we can treat
 the system  classically  considering  the equation of  motion  of the
 separation $r_{12}$ between two classical point charges (see appendix
 C).    Under     the   influence     of the     external    potential
 $v_{ext,1}=\omega_1(t)^2 r^2_{12}/2$ the    equilibrium    separation
 $r_{12}$ oscillates according to the classical equation of motion 
\begin{equation}{\ddot r_{12}}=E_{ext,1}(t)-3\omega_0^2(r_{12}-r_0)
\label{clar_0}\end{equation}
with $E_{ext,1}(t)=-\omega_1^2(t)r_0$ and   $r_0$ is  the  equilibrium
separation in  the absence of  the external  field  (we use the linear
response approximation).  We can now define the ``adiabatic external''
field as the one that produces the  same deviation from equilibrium as
$E_{ext,1}(t)$, under  static conditions  (${\ddot  r_{12}}=0$).  That
means that 
\begin{equation}E_{ext,1}^{ad}(t)=3\omega_0^2(r_{12}(t)-r_0)
\end{equation}
where $r_{12}(t)$ is the solution of (\ref{clar_0}).  From this we can
deduce that 
\begin{equation}{\dot v}_1= E_{ext,1}(t)-E_{ext,1}^{ad}(t)
\end{equation}
where $v_1$ is  the exact velocity  field in this limit.  This implies
that $E_{ext,1}^{dy}(t)=0$ in this limit. 

{\bf Non extreme cases} 
 
In this  case the problem  is to find  a  good approximation for ${\bf
E}_{ext}^{ad}(r,t)$.  In    the  case    we  are   considering   ${\bf
E}_{ext}(r,t)=-\omega (t)^2{\bf r}$ so  in order to  have a simple and
separable form for ${\bf E}_{ext}^{ad}(r,t)$, we can choose: 
\begin{eqnarray}{\bf E}_{ext}^{ad}(r,t)&\approx&-\alpha(t){\bf r} 
\label{vext_ad_apx}\\
{\bf E}_{ext}^{dy}(r,t)&\approx&  -(\omega  (t)^2{\bf r}-\alpha(t){\bf
r})\label{vext_na_apx}\end{eqnarray}   and  determine   $\alpha$    by
optimizing the density.

For the ``low  correlation'' parameter  $\omega_0=1$ the approximation
${\bf  E}_{ext}^{ad}(r,t)=-(\dot{\phi})^2{\bf  r}$  gives very    good
results and $\rho_0({\bf r};t)$ is indistinguishable from $\rho( r,t)$
within the numerical error.  The  results for the ``high correlation''
parameter $\omega_0\approx 0.02$ are less  good.  They can be improved
using  $\alpha(t)=\omega_0^2(1+\varepsilon\cos(\Omega t))$ and  tuning
the   parameter $\varepsilon$.   In      every case,  also   in  these
intermediate  cases  ${\bf  E}_{xc}^{dy}(r,t)\approx  0$   within  the
numerical error. 

We  conclude that  for this  particular system  the dynamical part of
${\bf   E}_{xc}(r,t)$ is almost  negligible.   However we caution that
this is at least partly  a special  feature  of the harmonic system
studied here (see discussion in the following section)
and should not  be uncritically generalized to other systems. 

\section{Discussion and Summary} 
The comparisons performed in this paper between the exact xc potential
of a two-electron   harmonic atom and  several approximate expressions
for  this   quantity, constitute  an  extremely   severe test  of  the
approximations in question.  Aside from the exchange-only OEP, all the
approximations considered are based  on the homogeneous electron  gas,
and,  therefore,  are  expected to  be   valid only  for systems whose
density  is  slowly varying   on  the scale    of  the local   average
inter-electron distance.  This condition is certainly not satisfied by
our model system  -- not in the  weak correlation regime, in which the
length    scale  of density  variation   coincides   with the  average
inter-electron  distance, and much   less  in the  strong  correlation
regime, in which the latter greatly exceeds the former. In this light,
the fact that  the ALDA and GK produce  potentials reasonably close to
the  exact ones, although  qualitatively  incorrect at  large distance
from the center, should be regarded as an  unexpected success of these
approximations.

Another surprising  result of our study  comes from  the separation of
${\bf E}_{xc}(r,t)$ into an  ``adiabatic'' and a  purely ``dynamical''
part.  The somewhat counterintuitive result is that, in  the case of a
time-dependent harmonic  external  potential,  the dynamical   part of
${\bf E}_{xc}$ is  zero in the limits  of weak  and strong correlation
and  almost  negligible in between. This   happens at frequencies well
above the first excitation  threshold,  where the density response  is
far  from  adiabatic. 
In the weak correlation regime, this result depends crucially on the
form of the wavefunction in a parabolic potential. Therefore we don't
expect the conclusion to be generalizable to other potentials.
In the strong correlation regime, however, the reduction of the 
dynamics to harmonic oscillations about a classical equilibrium 
configuration appears to be a generic feature of the many-electron 
system. It is this feature that leads to the vanishing  of 
${\bf E}_{xc}^{dy}(r,t)$ in this regime.

These results throw some light on the surprising ability of the ALDA
to give good results even outside its natural domain of validity
(low frequency regime): in a system in which the non-adiabatic
corrections are small, a static functional of the density (such as
the LDA xc potential) which works well in the static regime, is
expected to give a reasonable time-dependent potential upon 
replacement of the static density with the time-dependent one. 

The recently  introduced  VUC approximation, contains a  ``dynamical''
correction to  ALDA (see eq. (\ref{eqVUC}))  and, in the light  of the
exact   behavior of ${\bf  E}_{xc}^{dy}(r,t)$   just underlined, it is
interesting to notice that the ``dynamical'' part of  VUC is, for this
system, small, becoming almost negligible for strong correlation. 

In summary,   we  have found that, for    this particular system,  the
``dynamical'' part of ${\bf E}_{xc}$ is almost negligible,
 and the ALDA,
GK and VUC  approximations    work reasonably  well at   all  coupling
strengths  (although  the VUC  underestimates ${\bf  E}_{xc}(r,t)$ for
weak correlation).  The OEP, as expected,  is reasonable only for weak
correlation.  The  main discrepancies are found to  occur at small $r$
and at  large $r$ (except  for the  OEP that  has the exact asymptotic
behavior).  The question of whether these results are generalizable to
more complex systems remains open. 

We  acknowledge    support from NSF  Grant   No.  DMR-9706788 and from
Research  Board   Grant RB     96-071     from  the University      of
Missouri-Columbia.  We thank C. A.  Ullrich, S. Conti, R. Nifos\'i and
C. Filippi for useful discussions.

\appendix 
\section{}
Making the change  of variable  ${\bf R}=  2{\bf {\cal  R}}$, eq.
(\ref{CM1}) can be rewritten as: 
\begin{equation}\label{CM2}
(-\nabla_{\bf      R}^2+\frac{1}{4}\omega^2(t)      R^2)\Psi_{CM}({\bf
R},t)=i\frac{\partial}{\partial  t}\Psi_{CM}({\bf R},t).\end{equation}
Separating angular  and radial  coordinates as in eq.(\ref{separ}),
 we obtain  the radial
equation: 
\begin{equation}\label{CMr}
(-\frac{\partial^2}{\partial  R^2}-\frac{1}{R}\frac{\partial}{\partial
R}+\frac{1}{4}\omega^2(t)              R^2+\frac{m^2}{R^2})\chi_{n,m}(
R,t)=i\frac{\partial}{\partial      t}\chi_{n,m}(  R,t).\end{equation}
Inserting into eq. (\ref{CMr}) the guess 
\begin{equation} \label{guess} \chi_{n,m}( R,t)=A(t)R^m\exp(B(t)R^2)
L^m_n(C(t)R^2) 
\end{equation}
(a generalization of the corresponding static solution), we obtain the
following equations for the time dependent coefficients: 
\begin{eqnarray} & &\label{eq1} i\dot{C}+8BC+4C^2=0 \\ & &
i\frac{\dot{A}}{A}+4B+4mB-4nC=0          \label{eq2}\\         &     &
i\dot{B}+4B^2-\frac{1}{4}\omega^2(t)=0 \label{eq3} .    \end{eqnarray}
With the ansatz $B=(i/4)(\dot{X}/X)$, eq. (\ref{eq3}) becomes: 
\begin{equation} \ddot X=-\omega^2(t) X \label{cl_eqa} , 
\end{equation}
the classical  equation  of motion for an   harmonic  oscillator.  The
solution $X(t)$ can be written as 
\begin{equation}X(t)=|X(t)|e^{i\phi(t)}=X_{\Re}+iX_{\Im}\label{cl_eq2}.
\end{equation}
Since $\chi_{n,m}(  R,t)$  must not diverge  as $R\to\infty$,  we must
 impose  that    the  real    part   of  $B(t)=B_{\Re}+iB_{\Im}$    be
 negative. Using (\ref{cl_eq2}), $B(t)$ can be written as: 
\begin{eqnarray} B(t)&=&-\frac{1}{4}\frac{d\phi}{dt}+\frac{i}{4}
\frac{d\ln|X|}{dt}\\
&=&-\frac{W}{4|X|^2}+\frac{i}{4}\frac{d\ln|X|}{dt} 
\label{B},\end{eqnarray}                                         where
$W=\dot{X}_{\Im}X_{\Re}-\dot{X}_{\Re}X_{\Im}$ is a constant, being the
Wronskian of two solutions of eq.  (\ref{cl_eqa}).  In order to have a
normalizable wave  function not   identically zero,  we have  then  to
impose that $W>0$ or equivalently that $d\phi/dt>0$. 

Requiring that $C\in\Re$,  from the  real part of  eq.(\ref{eq1}) and
from eq.(\ref{B}) we get 
\begin{equation}C(t)=(1/2)(W/|X|^2)
\end{equation}
which also satisfies the
imaginary part  of eq.(\ref{eq1}).  Now  we can solve eq.  (\ref{eq2})
from which, integrating, we get: 
\begin{equation} 
A(t)=A(0)\cdot
\exp\{-(m+1)\ln\frac{X(t)}{X(0)}-i(2n+m+1)(\phi(t)-\phi(0))\}
.\end{equation} $A(0)$  is determined  by the  normalization condition
$\int_0^{\infty} |\chi_{nm}(R,t)|^2RdR=1$, 
\begin{equation}A(0)=\sqrt{\frac{n!}{2^m(n+m)!}\Big(\frac{d\phi}{dt}
\Big|_{t=0} \Big) ^{(m+1)}}. 
\end{equation}
Finally inserting  all  the   expressions  for  the  coefficients   in
eq. (\ref{guess}), after some algebra we get: 
\begin{eqnarray} \chi_{n,m}( R,t;X)&=&\sqrt{\frac{n!}{2^m(n+m)!}}
\Big(\frac{d\phi}{dt}                                            \Big)
^{\frac{m+1}{2}}\exp\Big(i(2n+m+1)(\phi(0)-\phi(t))\Big)\cdot
\nonumber\\                &                   &                   R^m
\exp\Big((-\frac{d\phi}{dt}+i\frac{d\ln|X|}{dt})\frac{R^2}{4}\Big)
L_n^m\Big(\frac{d\phi}{dt}\frac{R^2}{2} \Big).\label{gen_sol}
\end{eqnarray}

\section{}
The  solution  of static  radial equation for  the  RM  channel can be
written as: 
\begin{equation}R_{n,l}(\rho)=\frac{u(\rho)}{\sqrt{\frac{\rho}
{\sqrt{\omega_r}}}}\end{equation} with 
\begin{eqnarray}u(\rho)&=&e^{-\frac{1}{2}\rho^2}\rho^s\sum_{\nu=0}
^{\infty} 
a_{\nu}\rho^{\nu}     \label{u},   \\ \rho&\equiv&\sqrt{\omega_r}r  \\
\omega_r&\equiv&\frac{\omega_0}{2}                                  \\
s&=&\frac{1}{2}+\sqrt{l}.\end{eqnarray} The coefficients of the sum in
(\ref{u}) are related by: 
\begin{eqnarray}
& &(s(s+1)-l+\frac{1}{4})a_1=\frac{a_0}{\sqrt{\omega_r}}\\ &   &((\nu
+s)(\nu                       +s-1)-l+\frac{1}{4})a_{\nu}-\frac{a_{\nu
-1}}{\sqrt{\omega_r}}+
[\frac{\varepsilon_r}{\omega_r}-1-2(\nu-2+s)]a_{\nu-2}=0 
\end{eqnarray} where $\varepsilon_r$ is  the part of the energy coming
from   the RM  channel and $a_0$ is fixed by the normalization 
condition. Imposing the  conditions  $a_{n-1}\not=0$,
$a_n=0$, $a_{n+1}=0$, the sum in (\ref{u}) can  be made finite and the
coefficients $a_{\nu}$  calculated.    From these conditions  we  also
obtain an  expression for the energy $\varepsilon_r=\omega_r(2n+2s-1)$
and an expression (less straightforward) for $\omega_r$. 

In  our calculations we have  truncated  the sum  in eq. (\ref{u}) at
$n=2$  obtaining $\omega_0=1$   (weak  correlation   case) and  at 
 $n=5$
obtaining       $\omega_0=(25-3\sqrt{33})/328\approx0.02$      (strong
correlation case).

\section{}
In the limit  of strong correlation (eq.  (\ref{stro})) and linear 
response  regime the potential energy can  be expanded up to second
 order about  the classical solution  and we  can also approximate the
 momentum  ${\bf p}$  with    the radial component   ${\bf  p}_r\equiv
 -i(\partial/\partial r)\hat{r}$ since, in this  limit, the dynamic of
 the problem is  basically confined in the   $\hat{r}$ direction.  The
 Hamiltonian  of the relative motion problem  eq. (\ref{RM1}) can then
 be approximated as: 
\begin{equation} H_{eff}(t)=\frac{p^2_r}{2\mu}+\frac{\mu}{2}{\tilde 
\omega}^2(r-r_0)^2-\mu E_{ext,1}(t)(r-r_0)\label{approRM} 
\end{equation}
where $\mu=1/2 $ is the reduced mass, $r_0=(1/\mu\omega_0^2)^{1/3}$ is
the  classical separation between    electrons in the  linear  regime,
${\tilde   \omega}^2(t)=3\omega_0^2+\omega_1^2(t)$ and $E_{ext,1}(t)=-
\omega_1^2(t)r_0$ can be viewed as an ``external force''. If we define
$r_1=r-r_0$, the deviation from the classical equilibrium position, we
can use the change of variable $r_1= y-x_0(t)$ so that the Hamiltonian
becomes \cite{V95}: 
\begin{equation} H(t)=\frac{p^2_r}{2\mu}+\frac{\mu}{2}{\tilde 
\omega}^2y^2-\mu{\tilde    \omega}^2x_0(t)y-  \mu   \ddot{x}_0(t)y-\mu
E_{ext,1}(t)y  \end{equation}  where  we  have  dropped the irrelevant
terms depending on the time alone.  If we impose that 
\begin{equation} {\ddot x}_0=- {\tilde \omega}^2x_0-E_{ext,1}(t) 
\label{x_0ap}\end{equation}
we get 
\begin{equation} H(t)=\frac{p^2_r}{2\mu}+\frac{\mu}{2}{\tilde 
\omega}^2y^2\end{equation} and the  problem reduces to a 1-dimensional
harmonic oscillator and can be solved exactly  in a way similar to the
one shown for the 2-dimensional harmonic oscillator in Appendix A. The
general solution takes the form: 
\begin{equation}\Psi(y,t)=(\frac{1}{2^nn!\sqrt{\pi}})^{{\frac{1}{2}}}
(\frac{d{\tilde\phi}}{dt})^{\frac{1}{4}}
e^{\frac{i}{2}\frac{\dot{{\tilde         X}}}{{\tilde          X}}y^2}
e^{i(\frac{1}{2}+n)({\tilde          \phi}(0)-{\tilde       \phi}(t))}
H_n((\frac{d{\tilde\phi}}{dt})^{\frac{1}{2}}y) 
\label{RMstro1}\end{equation}
where  $H_n(x)$ are the Hermite polynomials  and ${\tilde  X(t)}$ is a
{\it complex} solution of the classical equation of motion 
\begin{eqnarray} \ddot {\tilde X}&=&-{\tilde\omega}^2(t) {\tilde X} 
\label{cl_eqRM} , 
\\ {\tilde   X}(t)&=&|{\tilde X}(t)|e^{i{\tilde\phi}(t)}\end{eqnarray}
with a     phase  ${\tilde   \phi(t)}$     satisfying   the  condition
$d{\tilde\phi}/dt>0 $.    The solution  of  the original  problem  eq.
(\ref{approRM}) with $n=0$ is therefore 
\begin{equation} \Psi(r,t)=\frac{1}{\pi^{\frac{1}{4}}}
(\frac{d{\tilde\phi}}{dt})^{\frac{1}{4}}        e^{\frac{i}{2}({\tilde
\phi}(0)-{\tilde                                            \phi}(t))}
e^{-i\mu\dot{x}_0(t)(r-r_0)}e^{\frac{i}{2}\mu       \frac{\dot{{\tilde
X}}}{{\tilde X}}(r-r_0+x_0(t))^2} e^{i\int_0^tdt'(\frac{\mu}{2}{\tilde
\omega}^2x_0^2-           \frac{1}{2}\mu\dot{x}_0^2)}\label{highwfapp}
.\end{equation}

We stress that  in the regime  of high correlation $\Lambda/r_0\to 0$,
where  $\Lambda\propto  \omega_0^{1/2}$   the width of   the  Gaussian
entering the   solution   (\ref{highwfapp}),  the wave   function   is
concentrated  around   $r_0$    (that justifies    the   approximation
(\ref{approRM}))  and  tends to   a $\delta$-function in the extreme
limit.

\newpage
\begin{table}[h]
\caption{Parameters used in our numerical calculations. 
}
\label{table1}

\begin{tabular}{rcccc}
~&$\omega_0$ &$  \Omega$ & $\Omega/ \omega_0$ & $\lambda$  \\ \hline  
 high   corr. &
$(25-3\sqrt{33})/328\approx0.02$ &0.1&$\approx 4.2$ & 0.1\\ low corr. 
 & 1 & 3.2 & 3.2 & 0.1 \\ 
\end{tabular}
\end{table}

\begin{figure}
\caption{
Lowest Floquet state electronic density  for the weak correlation case
($\omega_0=1$). The solid  line   is the exact static   result
while each of the  broken  lines corresponds  to different  times.
   In  the inset we    show   the corresponding velocity field
$v(r,t)$. Each solid line  corresponds to different times.}
\label{fig1}
\end{figure}

\begin{figure}
\caption{
Same as Fig. 1 but for the strong correlation case
 ($\omega_0\approx 0.02$).} 
\label{fig2}
\end{figure} 

\begin{figure}
\caption{ 
Exact  xc field   $E_{xc}(r,t)$  for    the weak correlation      case
($\omega_0=1$). The solid  line is the static  limit while each of the
broken   lines    corresponds   to     $E_{xc}(r,t)$    at   different
times.  Asymptotically $E_{xc}(r,t)\approx -1/r^2$,   as in the static
case. For comparison  the static LDA result is 
also plotted.} 
\label{fig3}
\end{figure}

\begin{figure}
\caption{ Same as Fig. 3 but for the strong correlation case
 ($\omega_0\approx 0.02$). 
} 
\label{fig4}
\end{figure}

\begin{figure}
\caption{  
Comparison   between  the  {\it    exact}   first  Fourier   component
$E_{xc}(r,\Omega)$  of the xc field and some of 
 its  most used approximations
 (weak correlation case, $\omega_0=1$).}
\label{fig5}
\end{figure}

\begin{figure}
\caption{ Same as Fig. 5 but for the strong correlation case
 ($\omega_0\approx 0.02$). 
} 
\label{fig6}
\end{figure}

\newpage
%******************************************************************************
\unitlength1cm 

\begin{picture}(20,20.0)
\put(-8.5,-13){
\makebox(10,15){
\includegraphics{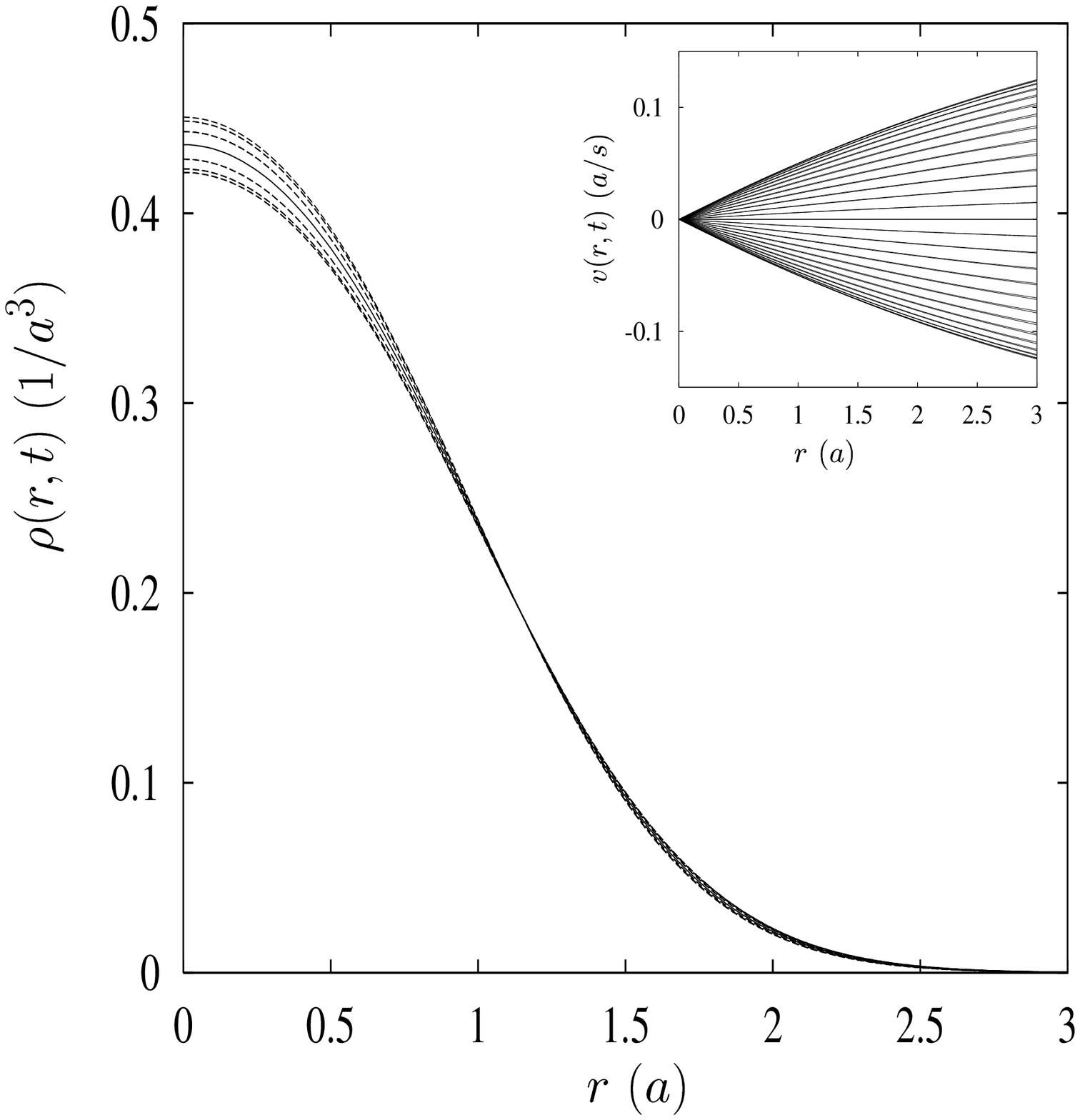}
}
}
\end{picture}

\newpage
%******************************************************************************
\unitlength1cm

\begin{picture}(20,20.0)
\put(-8.5,-13.0){
\makebox(10,15){
\includegraphics{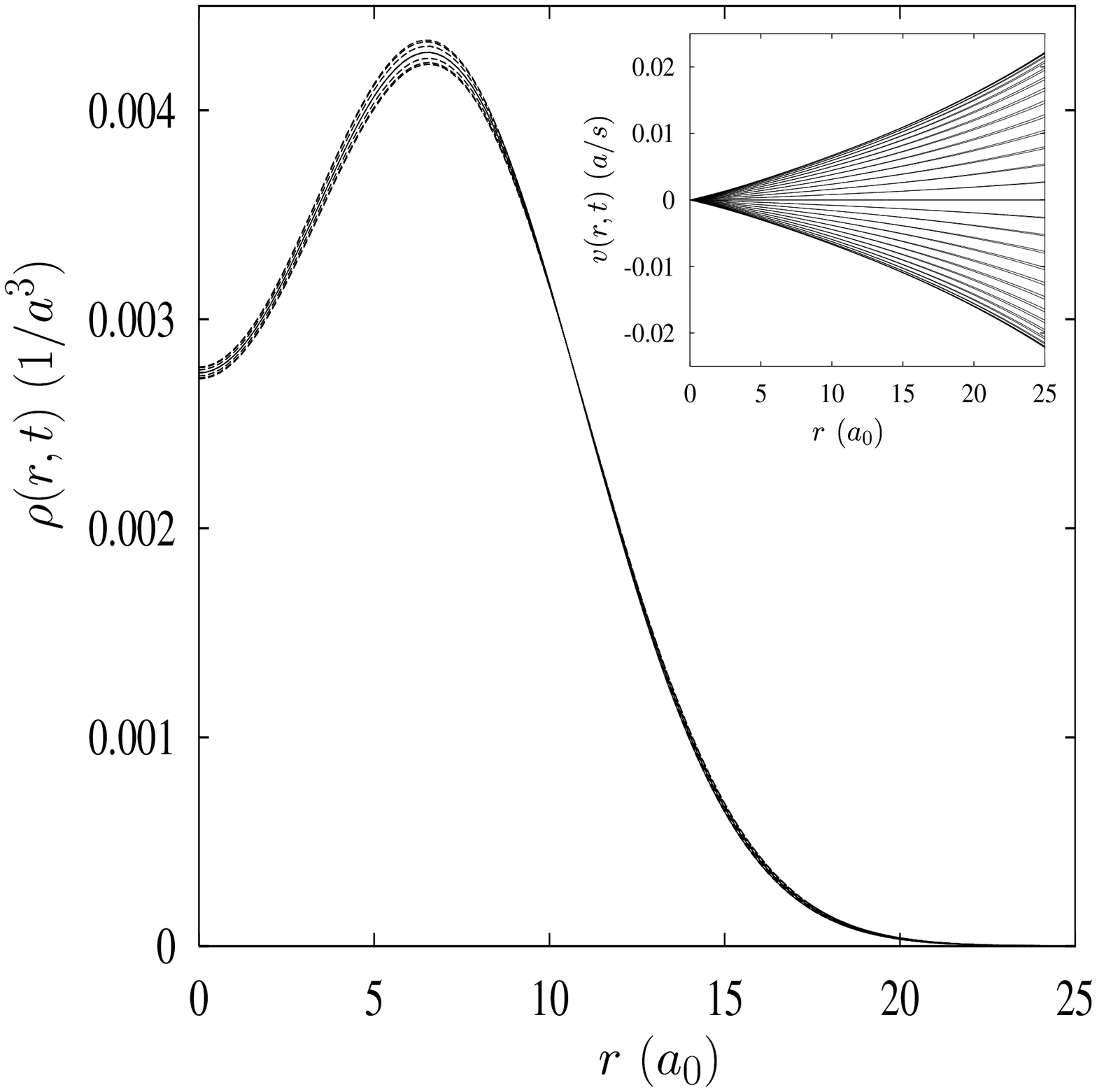}
}
}
\end{picture}
 
\newpage
%******************************************************************************
\unitlength1cm

\begin{picture}(20,20.0)
\put(-8.5,-13.0){
\makebox(10,15){
\includegraphics{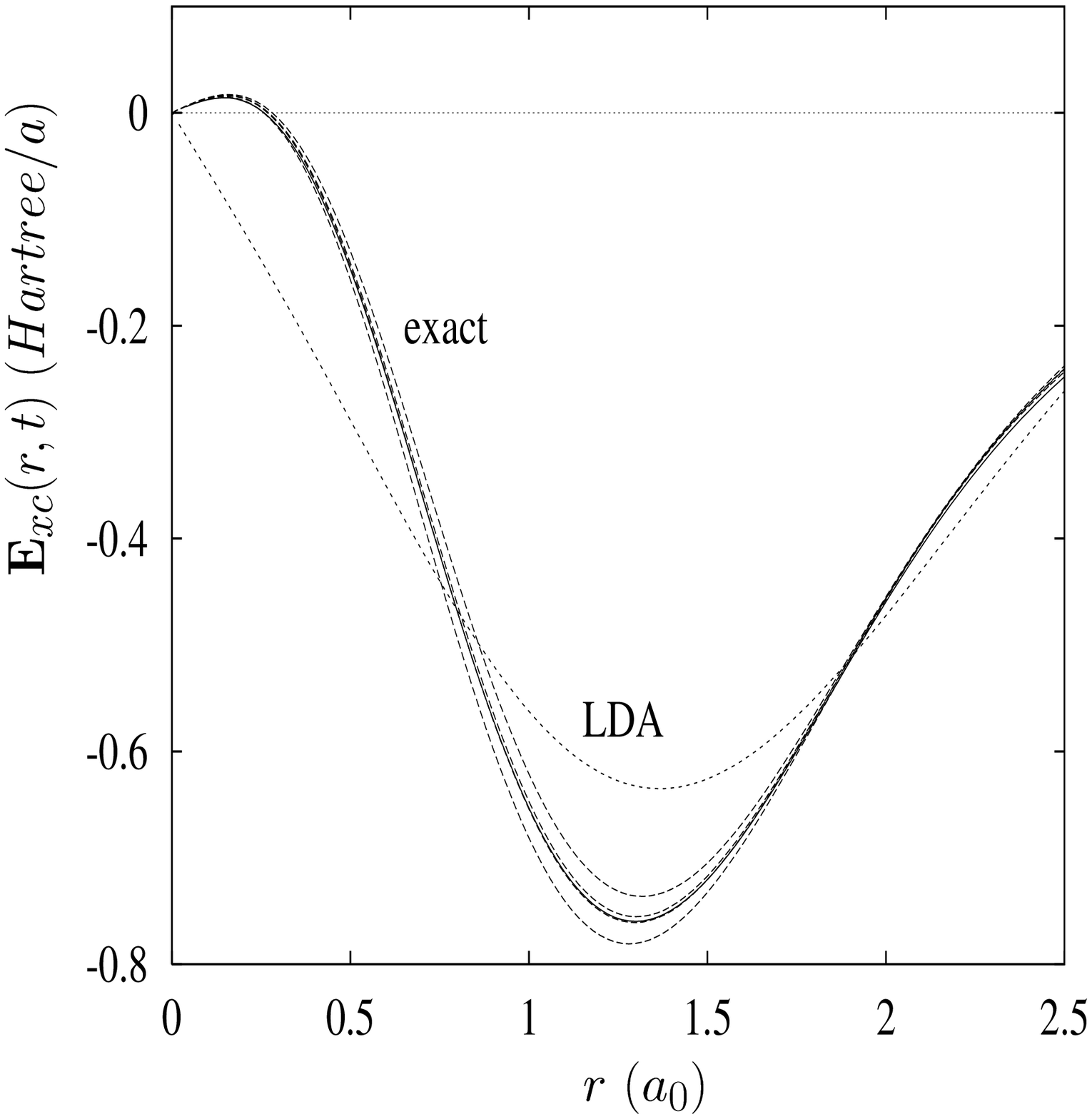}
}
}
\end{picture}

\newpage
%******************************************************************************
\unitlength1cm

\begin{picture}(20,20.0)
\put(-8.5,-13.0){
\makebox(10,15){
\includegraphics{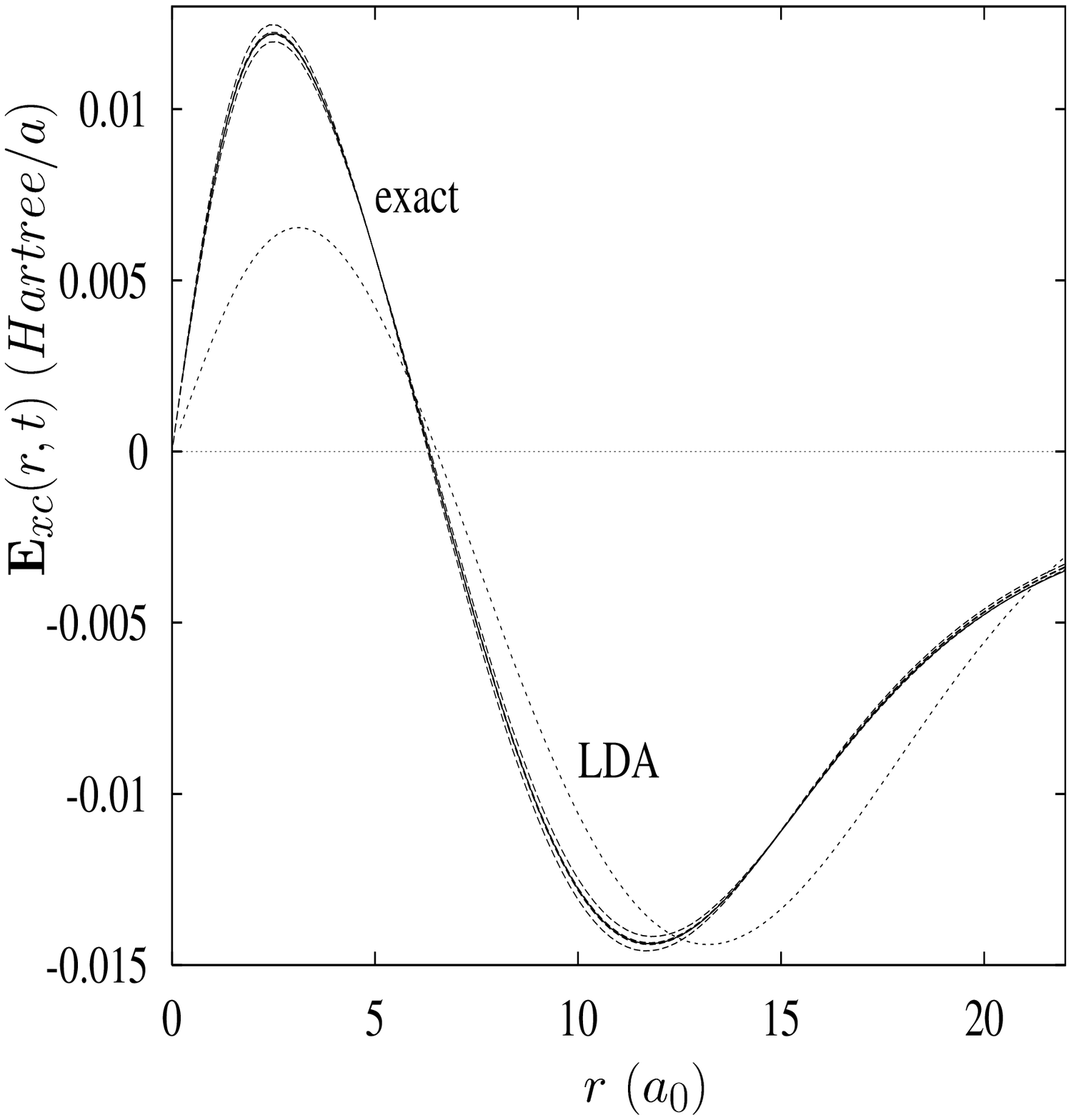}
}
}
\end{picture}
 
\newpage
%******************************************************************************
\unitlength1cm

\begin{picture}(20,20.0)
\put(-8.5,-13.0){
\makebox(10,15){
\includegraphics{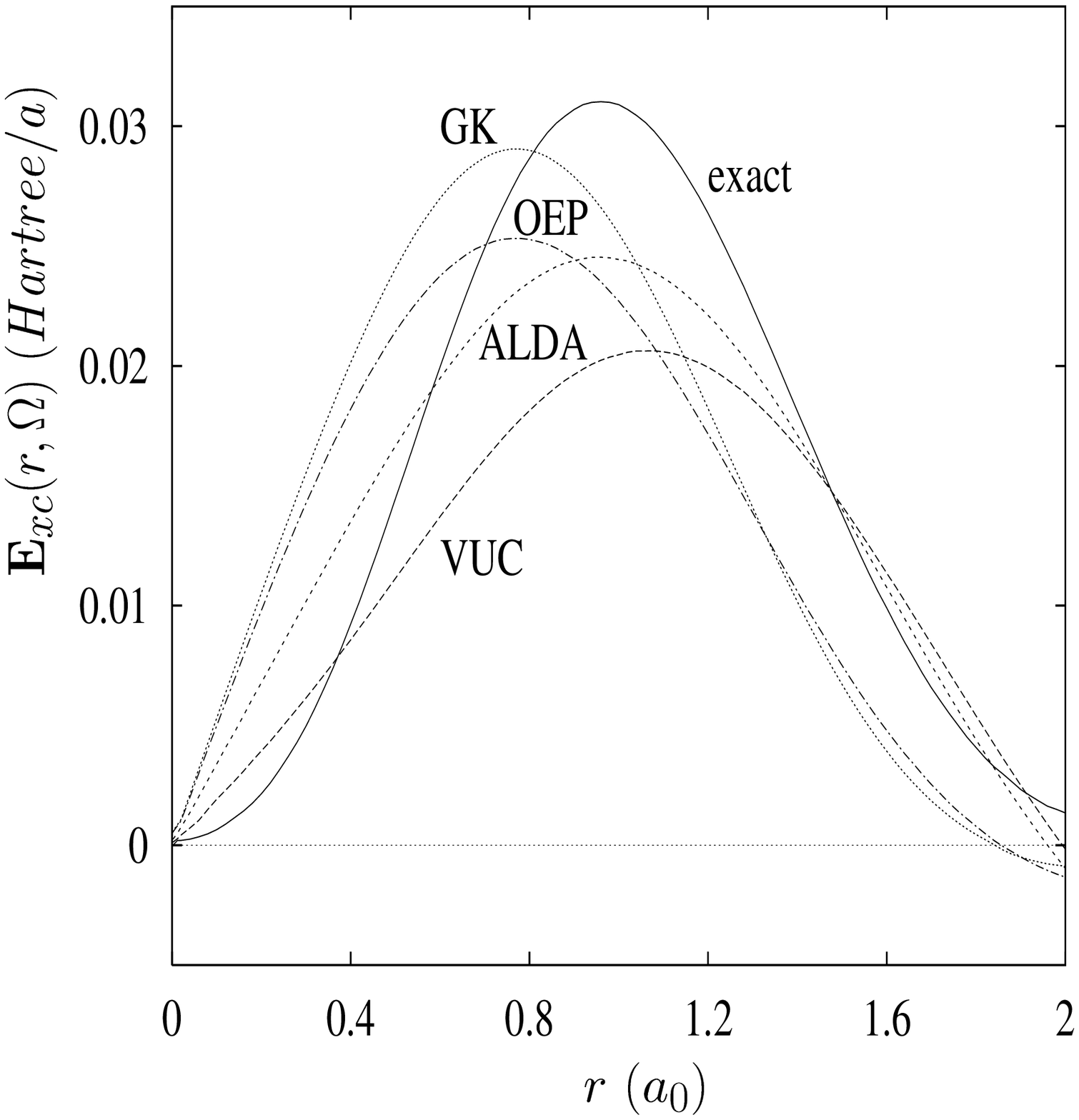}
}
}
\end{picture}

\newpage
%******************************************************************************
\unitlength1cm

\begin{picture}(20,20.0)
\put(-8.5,-13.0){
\makebox(10,15){
\includegraphics{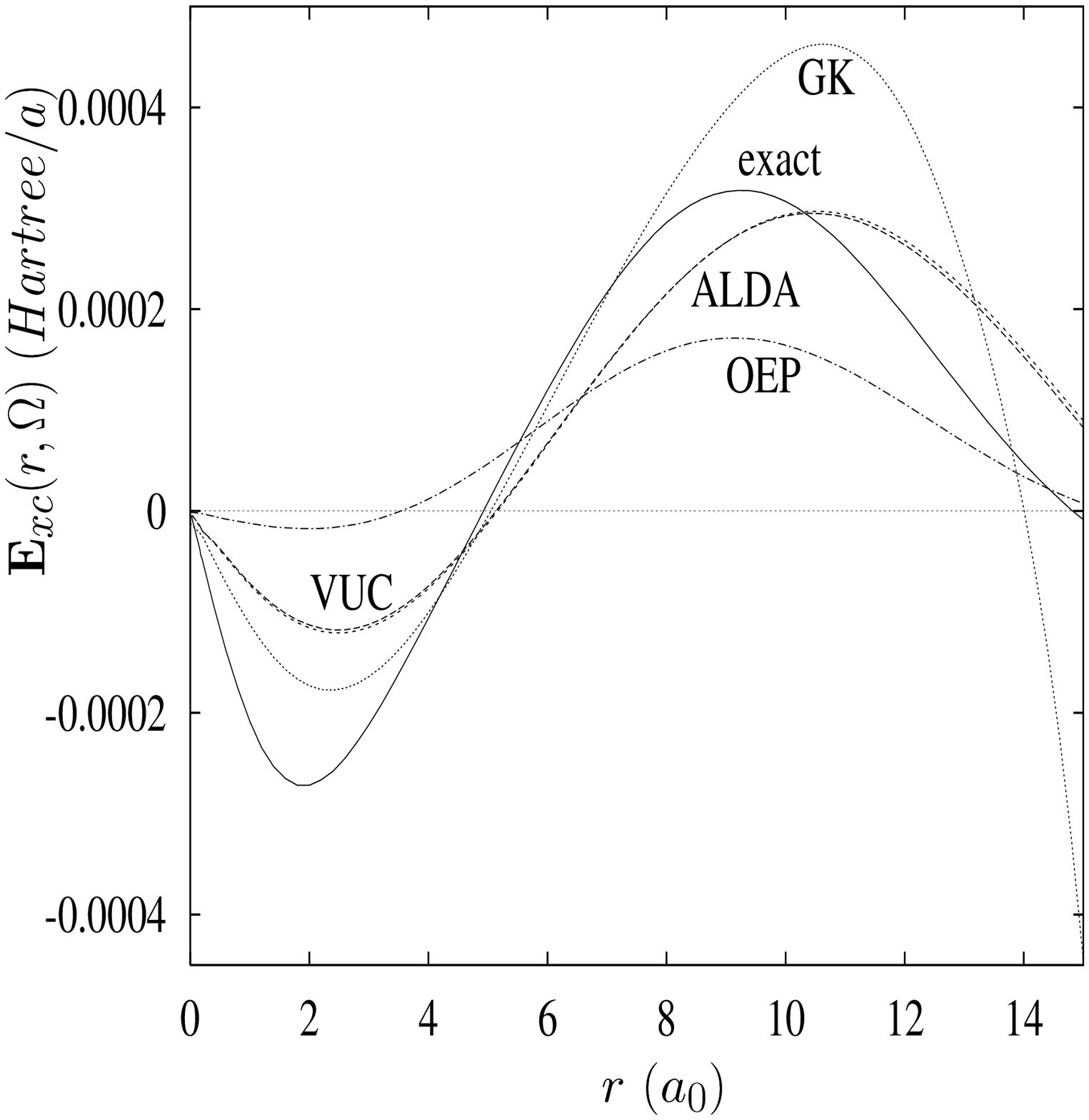}
}
}
\end{picture}

\end{document}